# A multi-layer agent-based model for the analysis of energy distribution networks in urban areas

Alberto Fichera[1], Alessandro Pluchino[2], Rosaria Volpe[1]*

[1]Department of Electrical, Electronics and Computer Engineering, viale Andrea Doria 6, 95125 Catania, Italy
[2]Department of Physics and Astronomy, via Santa Sofia 64, 95123 Catania, Italy

*rvolpe@dii.unict.it

**Abstract**

Significant research contributions and Directives approach the issue of the insertion of renewable-based energy systems on urban territory in order to face with the growing energy needs of citizens. The introduction of such systems gives raise to installers to both satisfy their energy demands and distribute eventual energy excesses to close neighbours. This paper presents a multi-layer agent-based computational model that simulates multiple event of the network of the energy distribution occurring within urban areas. The model runs on the NetLogo platform and aims at elaborating the most suitable strategy when dealing with the design of a network of energy distribution. Experimental data are discussed on the basis of two main scenarios within an operating period of 24 hours. Scenarios consider both the variation of the percentages of installers of renewable-based energy systems and the distance along which energy exchanges occur.

1. **Introduction**

The energy consumption in urban areas has significant implications for the transition towards environmentally sustainable cities [1], being the buildings sector acknowledged as one of the major contributor in $CO_2$ emissions. A way to face the growing carbon emissions of urban areas is recognized in the exploitation of renewable sources. As a consequence of the installation of renewable-based energy systems, such as photovoltaic panels, citizens shift from the condition of passive consumers to active producers [2], thus becoming able both to achieve the energy self-sufficiency and to exchange the own produced energy [3]. Indeed, the exchange of energy occurring from producer to consumer defines the basis for setting up a network of energy distribution and decreases the supply from fossil fuelled plants. Accordingly, the design of a network of energy distribution among consumers requires appropriate models that aim at assessing both the impact of renewable energy systems on the traditional supply and the connections in the resulting network to outline energy strategies.

The issue of the energy distribution among consumers and producers is treated in literature considering the optimization of the distributed technology from the economic and environmental viewpoint. For instance, the tool introduced in the work of Bracco et al. [4] optimizes the annual maintenance and operating costs of a distributed energy system, which provides heating, cooling and electricity to an urban neighbourhood. The paper of Mehleri et al. [5] pursues the same topic also exploring the optimal choice among several candidate distributed technologies and, in addition, optimizes the network of heat exchanges. Along with the minimization of the operating costs, the studies of both Alvarado [6] and Weber and Shah [7] address the reduction of the carbon emissions.

The cited studies are mainly technology driven, since they aim at selecting the optimal distributed technology that minimizes its overall operating costs or the emissions. Moreover, the exchanges occur within small-scale areas, not comparable to urban neighbourhoods. Therefore, despite the validity of the considered optimization methods, the increasing complexity of the energy exchanges configuration pushes towards approaches able to obtain solutions in shorter computational times and for growing complexity of the energy d configurations, such as those characterizing a neighbourhood . In this direction, agent-based models have been widely proposed as a valid technique to study energy systems characterized by interactions among the involved parts [8, 9], such as the interactions occurring due to the distribution of energy.

When attempting to provide a review of agent-based systems (ABSs) dealing with the exchange of energy, studies distinguishes from the role of agents in the distribution process. In the multi-agent model of Mbodji et al. [10] two agents aim at defining a management strategy to adapt the energy consumed to that supplied by renewable production sources of the system. In the paper of Sharma et al. [11] centralized agent guides all agents towards the balance of the demand for a peak shaving in a distribution system. The research of Bellekom et al. [12] explores the emerging rise of prosumers, namely consumers with a renewable energy production potential, and its implications for the grid management.

Moving forward, more detailed works also include financial issues in their studies. For instance, in Lopez-Rodriguez et al. [13], customers may contact act with brokering agents in order to participate in the market of the energy exchanges. In the agent-based model of Ye et al. [14], the energy distribution problem is formulated to admit autonomously negotiation among agents with the main objective to achieve efficient energy dispatch. Similarly, the paper of Kumar Nunna et al. [15] presents an agent based market model with price sensitive consumers.

On the same topic, but including a time-dependent analysis, the work of Degefa et al. [16] simulates the impact of prosumers agents minimizing their energy costs. The study of Misra et al. [17] analyses the energy trading problem with real-time demand estimation. A real-time control of the consumption and production of agents is also presented in the previous cited works [10, 11].

The main body of the listed literature in ABSs focuses on either the definition of management programs or financial aspects of the electricity exchanges. Nevertheless, although the issue of the energy exchanges has been widely considered, the energy distribution needs to be further deepened from a network perspective. Indeed, the energy exchanges occurring within the urban area configure a network of interactions among consumers that have installed renewable energy systems. Hence, to orient the design of a network of energy connections in the urban territory, appropriate models should examine the aspect of the energy distribution. The analysis of the network of energy distribution within a neighbourhood has been introduced in a previous work of the authors [18]. The model aims at both designing the optimal energy distribution network among consumers and minimizing the energy supply from the traditional fossil power plant. However, the study did not consider the variability of the energy demands and energy production during the day, which are instead included in this paper.

Therefore, this work defines a model that permits to determine practicable solutions for the design of the network of energy distribution within an urban area. To the purpose, the authors develop a multi-layer agent-based model able to simulate the network of the energy exchanges occurring among buildings equipped with autonomous energy production systems. The variability of energy demands and productions during the day are properly taken into account; indeed, the model considers the 24h energy cycle, which is relevant in case of a network of renewable sources.

2. **The agent-based model**

The installation of renewable based energy systems allows consumers to both reach the energy self-sufficiency and immediately distribute the eventual excess of produced energy. Considering each consumer as a node and each energy exchange as a link, the urban area may be modelled as a double-layer network [19], hereinafter called *energy distribution network* (see Fig.1), and simulated via an agent-based model. The main idea behind agent-based systems (ABSs) resides in the modelling of active entities, called *agent*, that interact in conformity to *rules* in order to achieve *tasks* within a defined simulation time. Each agent achieves the tasks through both its autonomous behaviour and the interactions with other agents.

In the elaborated model, the simulation time is denoted as $t = 0, \dots, T$ and two kinds of agents are introduced: on one side, *N* nodes-agents and, on the other side, one central-agent. Nodes-agents, hereinafter simply called agents, refer to the nodes of the energy distribution network that are characterized by an energy demand and may install renewable energy systems, whilst the central-agent is representative of the power plant, which provides for the traditional energy supply. As shown in Fig.1, in the top layer of the energy distribution network each node-agent equipped with renewable energy production (in green) interacts with all its neighbours (green or red) included in a circular area with a given radius, while in the bottom layer all the nodes-agents interact with the central-agent. Notice that in the top layer the central-agent is an isolated node and also some other node without its own energy production (in red) can be isolated.

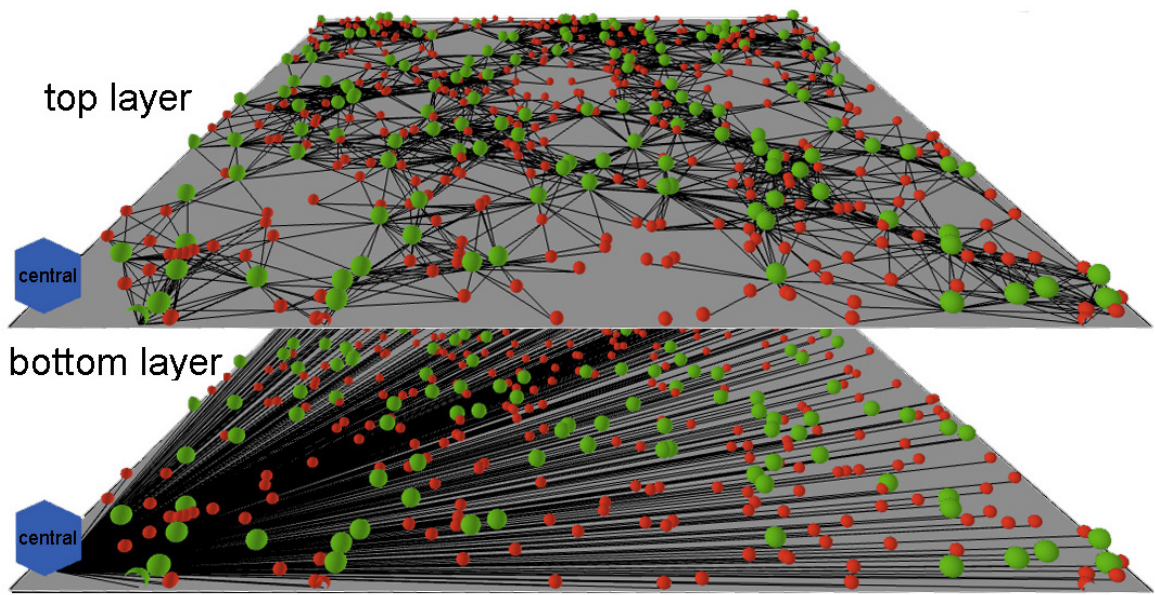

Fig.1 An example of the double-layer *energy distribution network*: in the top layer, the nodes-agents equipped with renewable energy production systems (called *producers*, in green) are connected with other producers or with non-producers nodes (in red) inside a given area; in the bottom layer, all the nodes are connected with the central-agent (in blue). Therefore, in total, *N* + 1 nodes are present in the network.

At each time $t$, the agent $i$ ($i = 0, \dots, N$) is characterized by an energy demand, $demand_{it}$, and an eventual energy production, $production_{it}$. Agents equipped with renewable energy production systems, hereinafter called *producers*, primarily satisfy their own energy demands by drawing energy from the own produced. If their demands have not been satisfied, the remaining gap is fulfilled from other producers and, at last instance, from the central-agent. Agents that are merely consumers receive energy either from producers or from the central-agent. Regarding the central-agent, the model assumes that its energy supply is unlimited, and, in particular, it corresponds to the energy demands not satisfied by the producers. The amount of energy that the central-agent supplies at time $t$ is defined as $central\_energy\_supply_t$.

As before introduced, agents take decisions to achieve the collective task of exchanging the produced renewable energy. They act according to a set of rules which depend on two main features: the energy status of each agent $i$ and the distance among agents.

The energy status of each agent i at time t is defined through the following variable:

$$energy\_surplus_{it} = production_{it} - demand_{it} \qquad (2.1)$$

If the $energy\_surplus_{it}$ assumes a positive value, the agent is able to distribute its exceeding energy to other agent of the network; on the contrary, a negative value means that the agent has not fulfilled its energy demand and requires energy either from other agents of the network or from the central-agent. Null values indicate that the agent either has used the entirely produced energy for the satisfaction of its own energy demand (if it is a producer) or has exactly fulfilled its energy demand by external sources. Of course, only producers can have positive values of energy surplus, while for normal agents this variable can be either negative or null.

The second important feature is the metrical distance among agents in the urban area. In fact, each agent can interact only with other agents present inside a given circular area, whose radius is fixed by a control parameter called $connection\_radius$. Actually, the top layer of the energy distribution network is built by randomly distribute the nodes-agents within the whole urban area and then by connecting producers nodes with all and only the other nodes located in the area limited by the connection_radius, which are considered as neighbours with whom to potentially exchange energy. In addition, as already said, in the bottom layer each agent is connected with the central-agent; this assumption is made to follow the actual configuration of the traditional energy grids. The set of agents and the set of links that connect agents into the two layers constitute the starting topology of the energy distribution network before each simulation run, i.e. at time $t = 0$.

Once the network of the feasible connections has been defined, the simulation starts and, at each time t > 0, agents behave in order to exchange the produced renewable energy and, consequently, decrease the energy supply from the central-agent. To properly describe such a dynamical process, the following three indexes are introduced. Each of them is evaluated for the entire time interval, i.e. from time $t = 0$ to time $t = T$.

The first index measures the fraction of links that agents really use for the exchange of energy, calculated with respect to the total number of links existing in the network; it is called $links\_percentage$ and is defined as

$$links\_percentage = \frac{active\_links}{active\_links + inactive\_links} \qquad (2.2)$$

where the variables $active\_links$ and $inactive\_links$ express, respectively, the links that are characterized by a number of energy exchanges which exceeds a given threshold (called *activation threshold*) during the total time interval $[0, T]$ and the number of the links that are not. In other words, setting up an activation threshold of 10%, for example, means that the model counts links as active if they are used *at least* the 10% of the total operating time. The evaluation of this index allows making assumptions on the effectiveness of the energy distribution network, in order to avoid unnecessary and costly interventions on territory. From its definition, it follows that the $links\_percentage$ varies within the interval [0,1]: the higher is its value, the more exploited is the energy distribution network.

The second index takes into account that the energy produced by the agents may be wasted, in the sense that there may be exceeding energy with respect to that needed by all the agents (included the producer) within the neighbourhood. This index is, therefore, influenced only by the producers that still have a positive energy surplus after the energy exchange and is indicated as

$$energy\_loss\_percentage = \frac{\sum_{t=0}^{T}\sum_{i=1}^{N}(production_{it} - demand_{it} - exchange_{it})}{\sum_{t=0}^{T}\sum_{i=1}^{N} production_{it}} \quad (2.3)$$

where $exchange_{it}$ is the amount of energy that agent i distributes to the set of its neighbouring agents at time t. More in detail, the (producer) agent i distributes its energy surplus to the set of neighbouring agents which require energy (i.e. with negative surplus) according to a priority list, in the way that agents with the smaller absolute value of surplus are supplied before the others. The $energy\_loss\_percentage$ may assume values within [0,1]. Of course, being renewable energy produced whatever the demand of the agents, it is preferable to have minimum values of energy loss. Therefore, this index is an operating indication aiming to measure the efficiency in the exploitation of the renewable energy system.

Finally, the third index estimates the percentage of energy supplied by the central-agent during the entire time interval and is indicated as

$$supply\_percentage = \frac{\sum_{t=0}^{T} central\_supply_t}{\sum_{t=0}^{T}\sum_{i=1}^{N} demand_{it}} \quad (2.4)$$

where the variable $central\_supply_t$ expresses the amount of energy that the central-agent supplies to all the nodes-agents at each time t, i.e.

$$central\_supply_t = \sum_{i=1}^{N} energy\_surplus_{it}, \quad \forall\, energy\_surplus_{it} < 0 \quad t = 0,\dots,T \quad (2.5)$$

As for the previous indexes, the values of the $supply\_percentage$ varies within the interval [0,1]: low values mean low supply, that is, the agents distribute a major amount of energy produced by means of renewable based systems.

In order to evaluate the best trade-off among the before introduced indexes, a further global measure with values in the interval [0,1] is introduced. This supplementary indicator is expressed as

$$index_{mix} = links\_percentage * (1 - energy\_loss\_percentage) * (1 - supply\_percentage) \quad (2.6)$$

The formulation of Eq.(2.6) considers that the $links\_percentage$ index is desirable to assume the highest value whilst the other two indexes, i.e. the $energy\_loss\_percentage$ and the $supply\_percentage$, are preferred to assume low values. To this purpose, they are considered as the complement to unity so that the ultimate goal of this study becomes that of maximizing this global index.

3. **Case study and discussion**

The introduced agent-based model aims at elaborating the most suitable strategy when dealing with the design of an energy distribution network among the buildings of an urban area. Buildings equipped with renewable based energy systems acquire the opportunity to exchange the own produced energy, beyond the chance to satisfy their demands. With the objective of designing an energy distribution network, it is essential to determine both which buildings may install renewable energy systems and which buildings to be connected for the exchange. To this end, a theoretical application is proposed to test the model and, precisely, an urban territory of $1300\,m \times 1300\,m$ with $N = 1000$ randomly placed nodes-agents is considered. Agents are representative of buildings characterized by an energy demand and a potential energy production deriving from the installed renewable energy systems. Simulations run on the NetLogo platform [20].

The main idea of the simulations is to reproduce the daily exchange of electricity among agents, in their capacity as consumers or producers.

Each agent corresponds to a building with 10 apartments and is characterized by the electricity demand profile of Fig.2. The electricity profiles are related the average household size and average electricity demand pro capita defined in the Eurostat 2016 [21]. Household demand covers the use of electricity for space and water heating and all electrical appliances. As regards to the electricity production, agents equipped with a photovoltaic panels exhibit an electricity production profile as the one in Fig.3, also obtained from the Eurostat statistics of 2016 [22]. In the intent of the present case study, two different electricity consumption and production are considered. Precisely, in the first scenario each agent (both consumer and producer) is characterized by the same electricity consumption profile of Fig.2 and each producer by the electricity production profile of Fig.3. This scenario is hereinafter called *constant profiles scenario*. In the second scenario, called *variable profiles scenario*, each agent varies its electricity production and consumption still according to the profiles of Figures 2 and 3, but also agreeing to a random uniform distribution with mean equal to the hourly value of the electricity consumption or production profile and standard deviation equal to one-third of the mean. Agents may exchange electricity within the neighbourhood defined by means of the chosen *connection_radius*. Each agent, during the time simulation, searches for other agents located within the defined neighbourhood and takes action to distribute its electricity surplus.

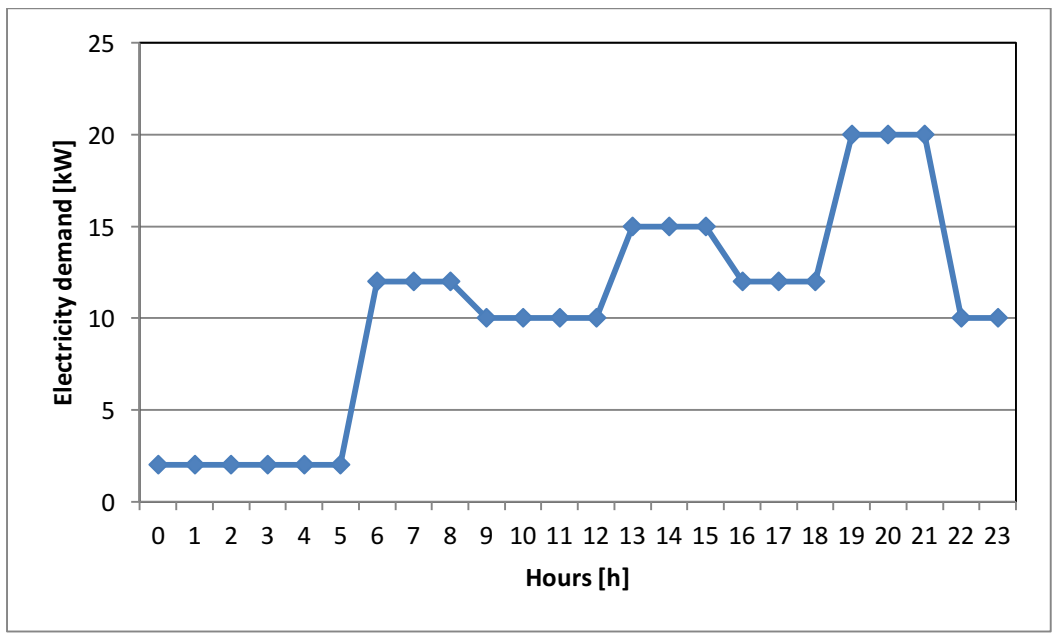

Fig.2 Electricity demand profile of each agent

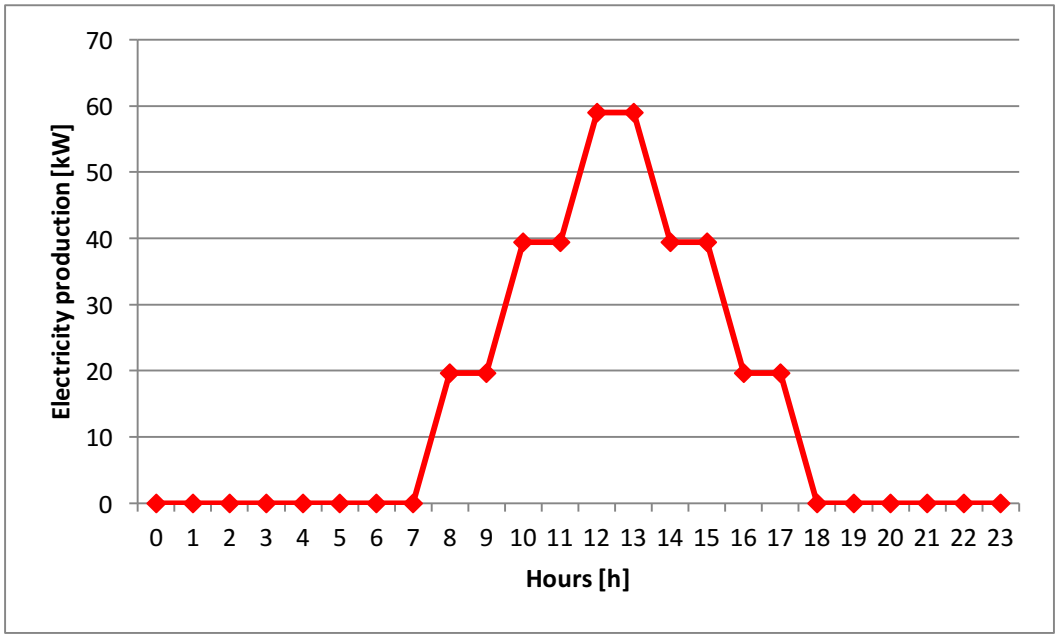

Fig.3 Electricity production profile of each agent equipped with a photovoltaic energy production system

The model has been implemented by considering the two following variables: the percentage of agents equipped with electricity production systems, briefly named *producers*, and the distance among agents.

Simulations run for both the constant profiles and the variable profiles scenarios at different percentages of producers and distances. Moreover, each scenario is also analysed for different links activation thresholds during the time period. In particular, the chosen threshold values correspond to the percentages of 0%, 5% and 10%.

All simulations are conducted for the 24 h cycle; however, since the electricity production of agents occurs in the time period from 8:00 to 18:00, the $links\_percentage$ index for every links activation threshold, the $energy\_loss\_percentage$ index and the $supply\_percentage$ index are here reported in relation to this specific time period. Indeed, since the time slots from 0:00 to 8:00 and from 18:00 to 24:00 are not characterized by renewable solar production, in these periods the demands of agents are solely satisfied from the central-agent, as in the traditional grid.

With the aim of illustrating the functioning of the model, a T=24 h simulation is carried on. To the purpose, the electricity distribution within the introduced hypothetical area populated by $N = 1000$ agents is reproduced for a fixed distance of connection, chosen as $connection\_radius = 150\ m$, and for a percentage of 30% of agents that have installed photovoltaic panels. Results are briefly summarized through the graphical interface of NetLogo in Fig.4.

Fig.4 reports the NetLogo environment at the end of a 24 h (i.e. 1440 min) simulation. The map displays both green and red nodes, representing respectively buildings that are responsible for the exchange or not, together with the corresponding links of the top layer. Counters for nodes, activated and inactivated links, time and demand are shown on the left of the interface. In the central part of the interface, three plots show the behaviour of the main indexes ($links\_percentage$, $supply\_percentage$ and $energy\_loss\_percentage$) as function of time, with a print time-step (called *supply_time*) of 15 minutes.

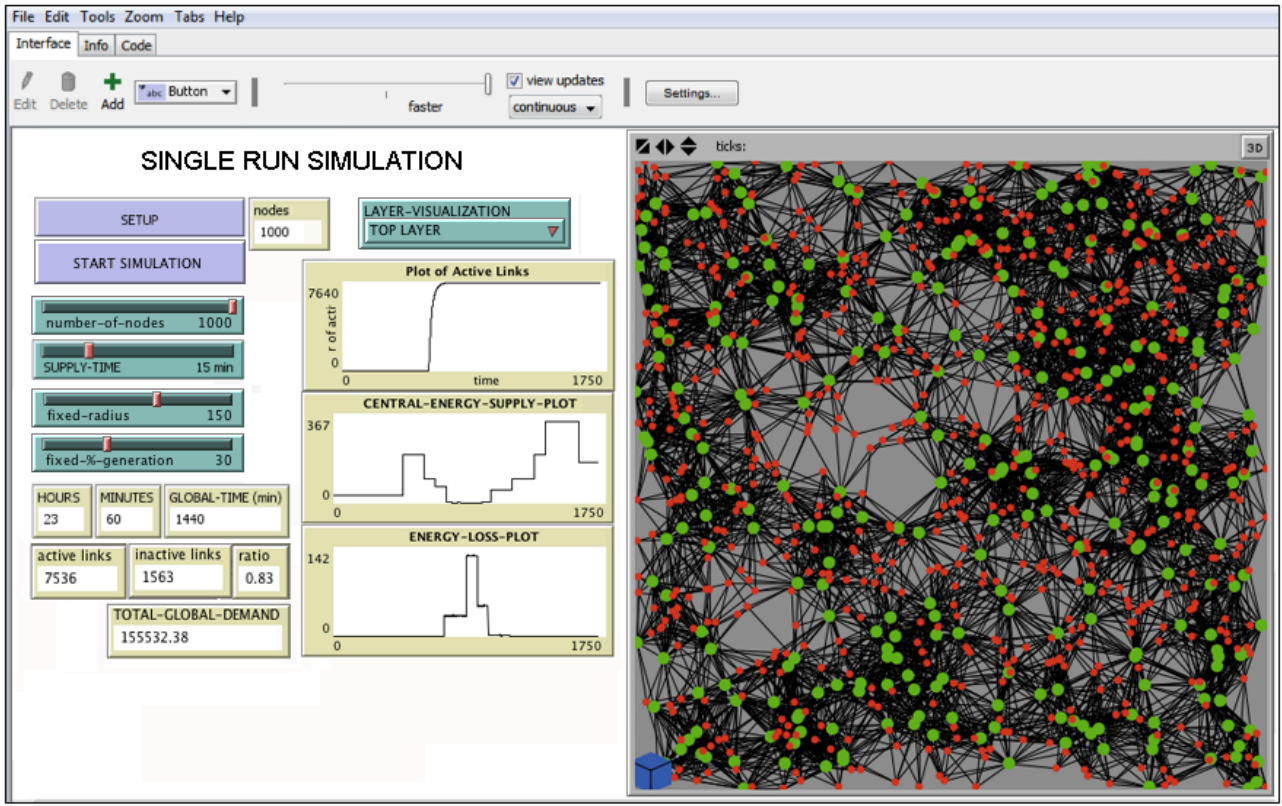

Fig.4 Graphical interface of NetLogo for a 24 h simulation, a $connection\_radius = 150\ m$, and for a percentage of 30% of producers and for the 0% of the links activation threshold

In particular, it results that: the links activation starts at 8:00 in the morning and rapidly reaches its asymptotic value (83%); the supply from the central-node starts to increase from its low night level around 6:00 and stays quite high for a couple of hours until, at 8:00, the renewable energy production becomes effective: then, as expected, this supply decreases towards zero due to the local energy exchange but, after 12:00, it increases again until it reaches its maximum value during the evening hours; consequently, the profile of the energy loss is concentrated in the central part of the day and reaches a maximum around 12:00. Finally, by integrating the last two curves over time and using Eq.(2.6) one could extract the final value of the global $index_{mix}$. In the next two paragraphs the behaviour of all these quantities will be studied as function of either the connection radius or the percentage of producers and within two distinct scenarios, considering first a constant profile of energy consumption and production, then a variable one.

*3.1 Scenario 1: constant profiles of both electricity consumption and production*

In the first set of simulations, the electricity consumption profile is equal for all agents of the network and corresponds to the profile of Fig.2. Similarly, agents equipped with photovoltaic panels display the electricity production profile of Fig.3 with no distinctions. The simulations are performed, on each occasion, for fixed values of the $connection\_radius$, i.e. the maximum distance along which agents may exchange energy, and considering different percentages of producers. The fixed values of the $connection\_radius$ are reported in Table 1, whilst the percentage of producers varies from 0% to the 100%. The increase in the values of the $connection\_radius$ indicates that at each simulation step a larger area is gradually considered and, therefore, a major number of connections are established. The percentages of producers are related to the total amount of agents of the network. In particular, increasing the percentage of producers implies that a major amount of agents installs photovoltaic panels.

| $connection\_radius$ [m] |
|---|
| 50 |
| 100 |
| 150 |
| 200 |

Table 1. Values of the $connection\_radius$

For each set of simulations, the trends of the $links\_percentage$ index, the $energy\_loss\_percentage$ index and the $supply\_percentage$ index at different links activation thresholds are recorded. The graphs of the $links\_percentage$ index are the only reported graphs that contain the specification of the considered threshold. Regarding both the $energy\_loss\_percentage$ and the $supply\_percentage$ indexes, instead, a unique graph is representative for all the thresholds. As a reminder, this choice derives from the evidence that the links activation threshold influences only the $links\_percentage$ index, since it corresponds to the calculation of how many times links are used for the energy distribution and does not account for electricity quantities.

The links activation threshold is initially set to 0%, i.e. each link is counted as active if it is characterized by *at least* one electricity exchange during the time period $[08:00, 18:00]$. Subsequently, the threshold is enlarged to the 5% and 10%. The trends of the $links\_percentage$ index for the three chosen links activation thresholds are reported in Fig.5.

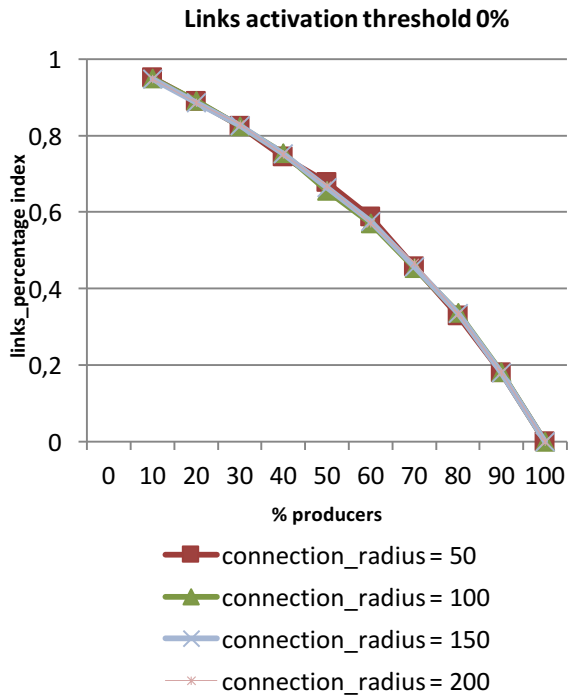
(a)

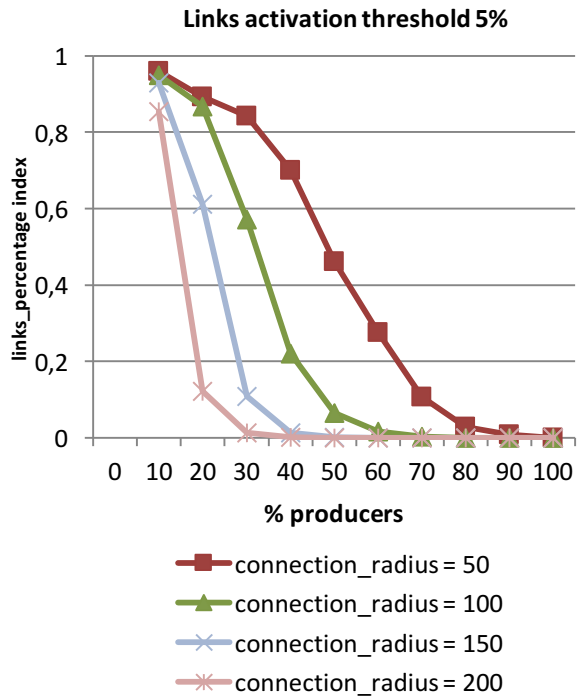
(b)

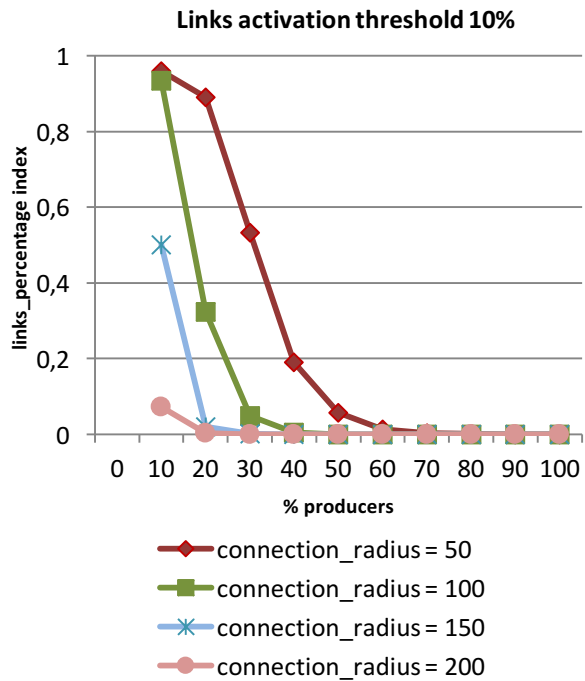

(c)

Fig.5 Trends of the *links_percentage* index in correspondence to: (a) links activation threshold 0%; (b) links activation threshold 5%; (c) links activation threshold 10%

The *links_percentage* index decreases as a function of the percentage of producers until the nil value in correspondence of the 100% of agents; i.e. the electricity produced by each agent is used for own purposes and, therefore, not distributed. As a result, not any link of the distribution network is used and the *links_percentage* index is null. Vice versa, the maximum values of the *links_percentage* index are recorded in correspondence of small percentages of producers. Into detail, the consideration of the links activation threshold significantly discriminates the results. Accordingly, when threshold is 0% (Fig.5a), the trends of the *links_percentage* index are similar independently of the *connection_radius* values. In these cases, the maximum value of the index is yielded in correspondence with the 10% of producers. The insertion of a links activation threshold greater than 0% (Fig.5b and Fig.5c), instead, returns a reduction of the *links_percentage* index at increasing the *connection_radius*; this is even truer for the threshold 10% in Fig.5c. This means that the majority of the links of the network is used less than the 5% or the 10% of the entire operating time interval. However, as a general result, the percentage of 10% of producers guarantees the maximum exploitation of the links of the distribution network. This is all the more true when distances equal to $connection\_radius = 50$ or $connection\_radius = 100$ are considered and for which the *links_percentage* index is almost near to the 0.93, i.e. the 93% of the total links established in the initial topology of the network are characterized by an electricity exchange.

The following graphs of Fig.6 and Fig.7, respectively representing the *energy_loss_percentage* index and the *supply_percentage* index, are reported by way of example for the threshold 0%. The *energy_loss_percentage* index rises when increasing the percentage of producers. In point of fact, a major electricity production yields a major surplus compared to the demands of the agents in the neighbourhood, thus resulting in an increase of the electricity that is not immediately distributed and, therefore, "lost". Concerning the distance of connection, instead, the amount of electricity that is considered as "lost" is slightly greater for $connection\_radius = 50$ compared to the higher distances.

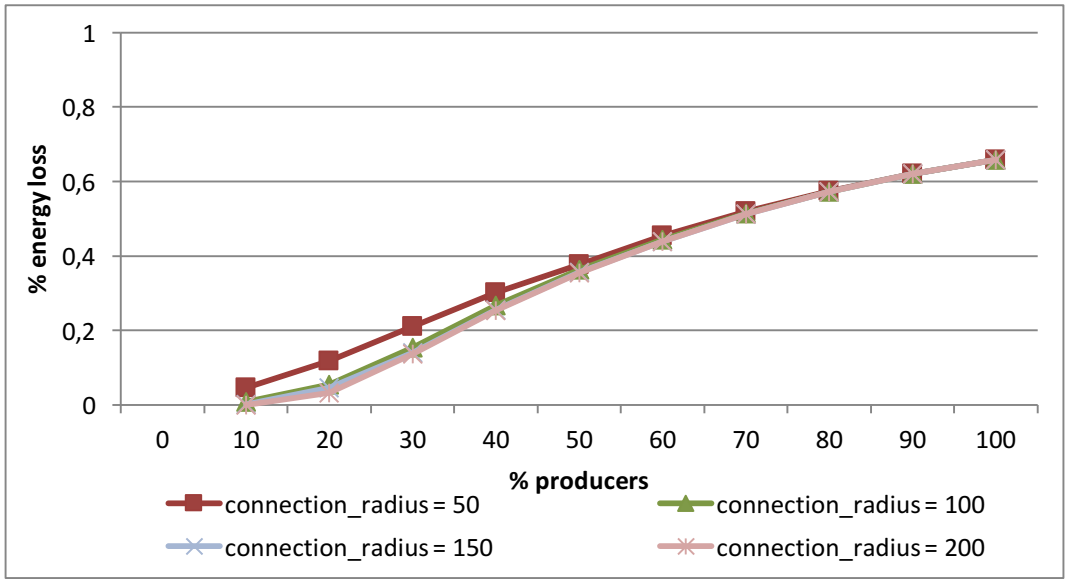

Fig.6 The $energy\_loss\_percentage$ index at the different $connection\_radius$ values

Finally, the $supply\_percentage$ index in Fig.7 diminishes at increasing percentages of producers. This is manifested in the fact that the increase of the installation of renewable systems allows agents to achieve the energy self-sufficiency, thus reducing the supply from the central-agent. Anyway, the central-agent keeps its significant role in the time interval occurring between 18:00 and 8:00, when the photovoltaic panels have a nil production.

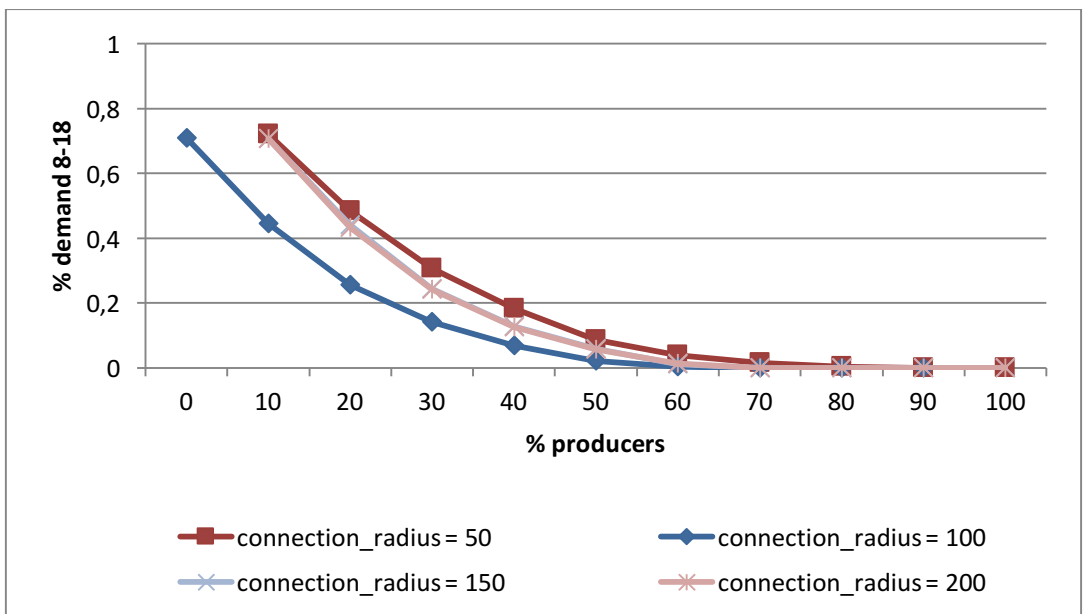

Fig.7 The $supply\_percentage$ index at the different $connection\_radius$ values

As concluding remark, the $index_{mix}$ reporting the best trade-off among the indexes is considered and shown in Fig.8.

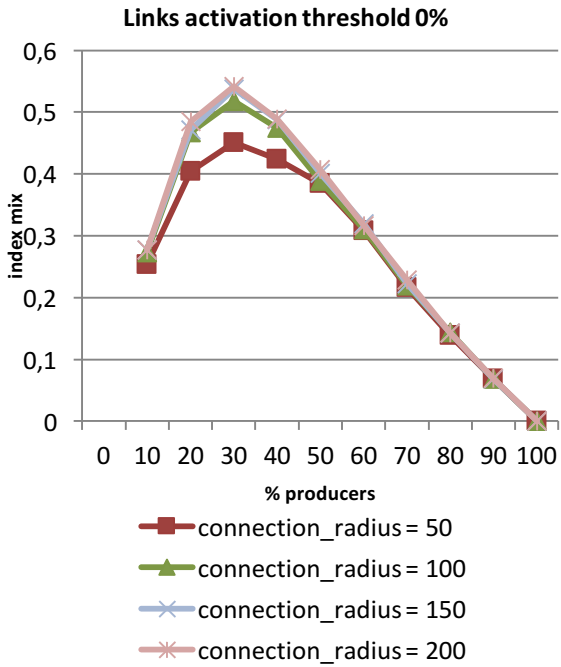
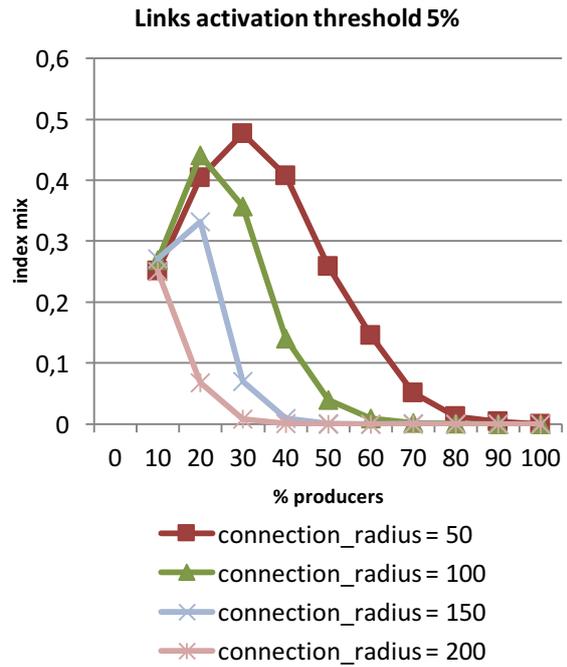
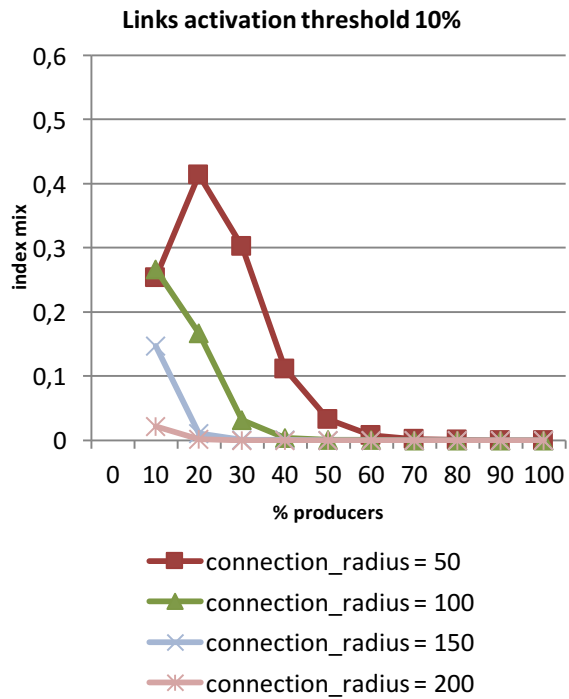

Fig.8 Trends of the $index_{mix}$ in correspondence to: (a) links activation threshold 0%; (b) links activation threshold 5%; (c) links activation threshold 10%

The examination of the $index_{mix}$ allows inferring conclusions about the relationship between the thresholds and both the distance of connection and the percentage of producers. It comes clearly out how the higher values of the $connection\_radius$ are preferable when no restrictions about the links activation are assumed. Vice versa, when considering thresholds greater than the 0%, the value of $connection\_radius = 50\ m$ returns the maximum value of the $index_{mix}$ and, as a confirmation, at increasing distances, the $index_{mix}$ decreases. This result means that the electricity distribution network performs more efficiently, i.e. with major activation of the links and with both minor electricity loss and demand to the central agent, when agents act in a spatial limited neighbourhood. Moreover, the percentage of producers that guarantees the best performance is around the 20% and the 30%.

The trends of the $index_{mix}$ for the different thresholds at fixed values of the $connection\_radius$ are compared in Fig.9.

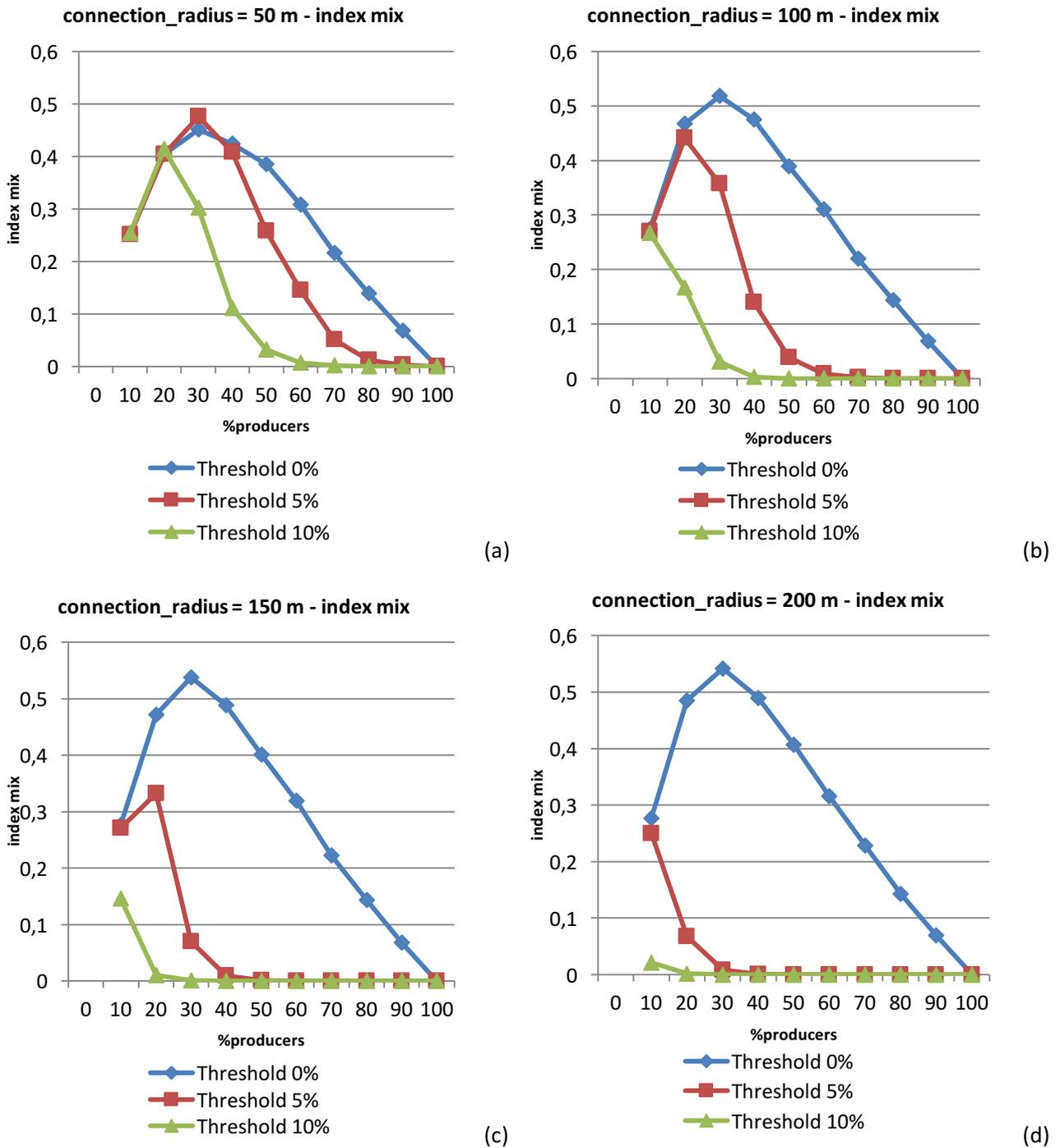

Fig.9 Threshold comparison for the $index_{mix}$ at: (a) $connection\_radius = 50\ m$; (b) $connection\_radius = 100\ m$; (c) $connection\_radius = 150\ m$; (d) $connection\_radius = 200\ m$

In Fig.9a the insertion of the links activation thresholds causes a decrease of the $index_{mix}$ for fixed percentages of producers are considered (excluding the percentages of 10% and 20% for which the curves are almost identical). At increasing values of the $connection\_radius$, the diversity among the thresholds is even more evident. In particular, in the case of 200 m (Fig.9d), the $index_{mix}$ is about null when threshold 10% is applied.

*3.2 Scenario 2: variable profiles of both electricity consumption and production*

The second set of simulations runs considering variable profiles of both electricity consumption and production for fixed values of the $connection\_radius$, by varying the percentages of producers. The values of the $connection\_radius$ are the same applied to the precedent scenario and reported in Table 1. Similarly, the percentage of producers varies from 0% to the 100% and the links activation is set to the 0, 5 and 10% in each simulation step. The $links\_percentage$ indexes are reported in Fig.10.

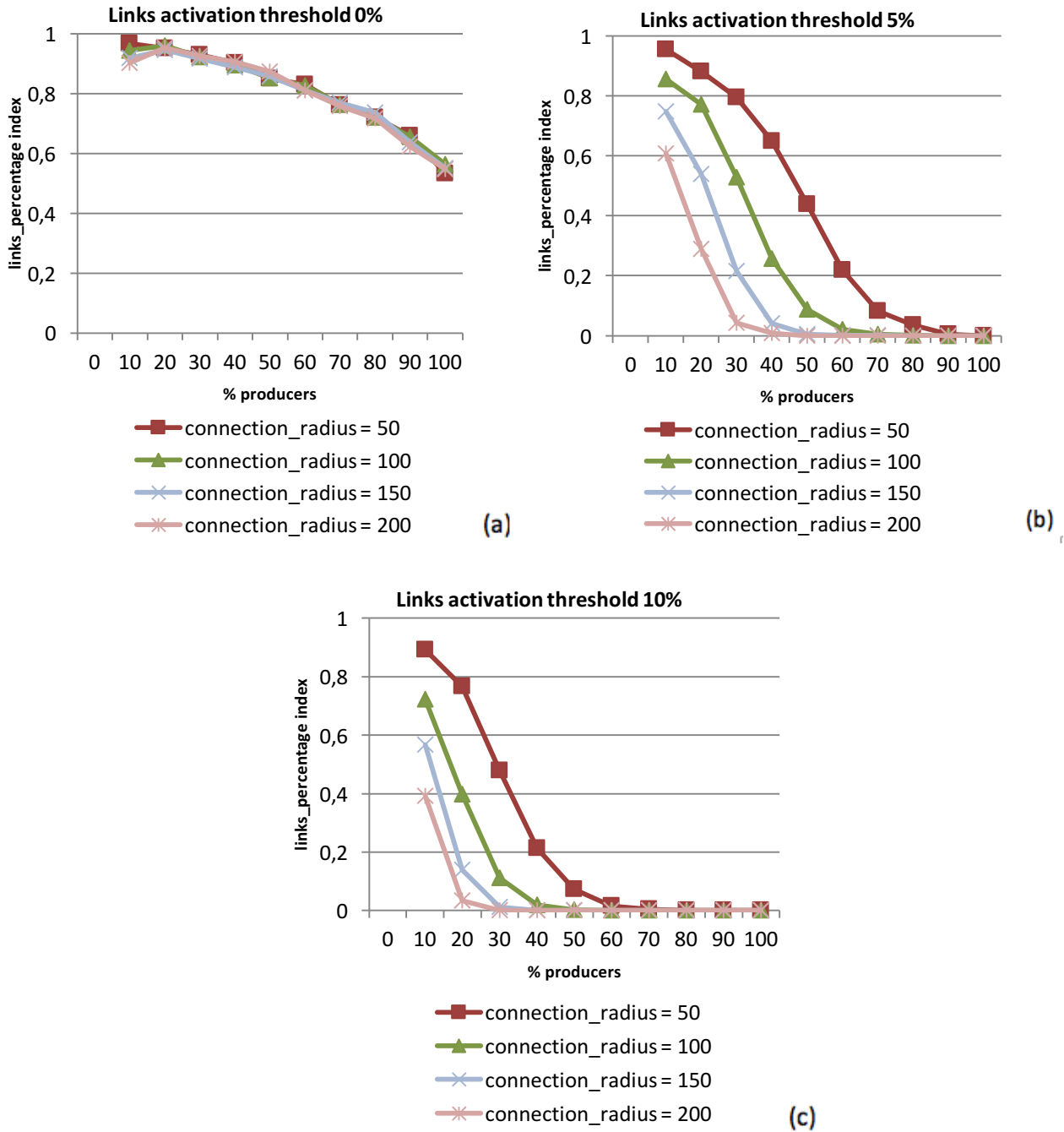

Fig.10 Trends of the $links\_percentage$ index in correspondence to: (a) links activation threshold 0%; (b) links activation threshold 5%; (c) links activation threshold 10%

The *links_percentage* index when threshold 0% is considered, i.e. Fig. 10 (a), displays similar trends at increasing values of the *connection_radius*. Moreover, its trend indicates how low percentages of producers ensure a higher exploitation of the links of the distribution network. Differently from the previous scenario (Fig.5a), the *links_percentage* index does not assume nil values due to the fact that the electricity demands and productions of the agents are here variable. Instead, being scenario 1 characterized by constant demands and productions, all agents are producers and consumers in the same way; therefore, given that the 100% of the agents are producers, no electricity exchange is feasible. When thresholds are included in the simulations (Fig.10b and Fig.10c) the maximum values of the *links_percentage* index are recorded in correspondence to a distance equal to $connection\_radius = 50\ m$, and, anyway, for the 10% of producers. Moreover, compared to the previous scenario (Fig.5b and Fig.5c), the *links_percentage* index decreases more uniformly at increasing the *connection_radius*. Or rather, a minor number of links are used for the distribution although the permitted distance of connection is higher.

The *energy_loss_percentage* index and the *supply_percentage* index for the threshold example 0% are reported in Fig.11 and Fig.12.

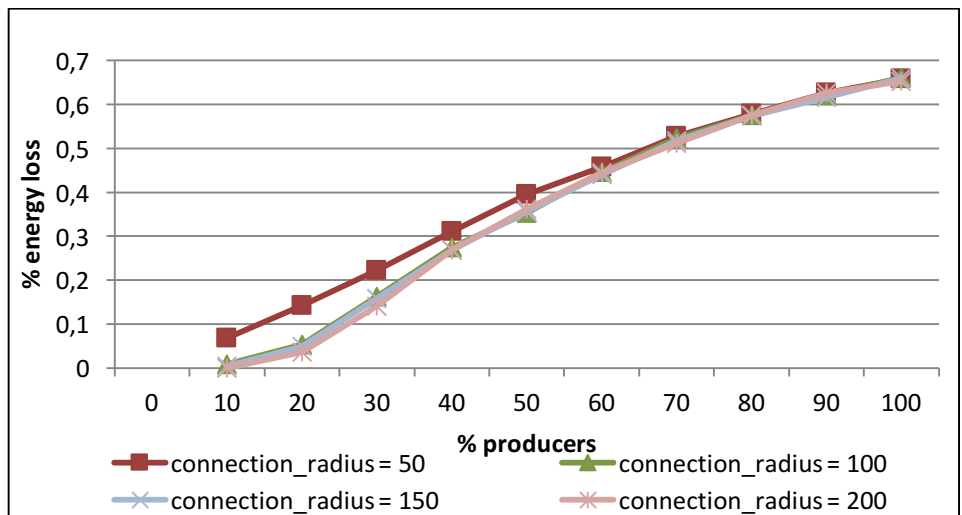

Fig.11 Trends of the *energy_loss_percentage* index in correspondence to links activation threshold 0%

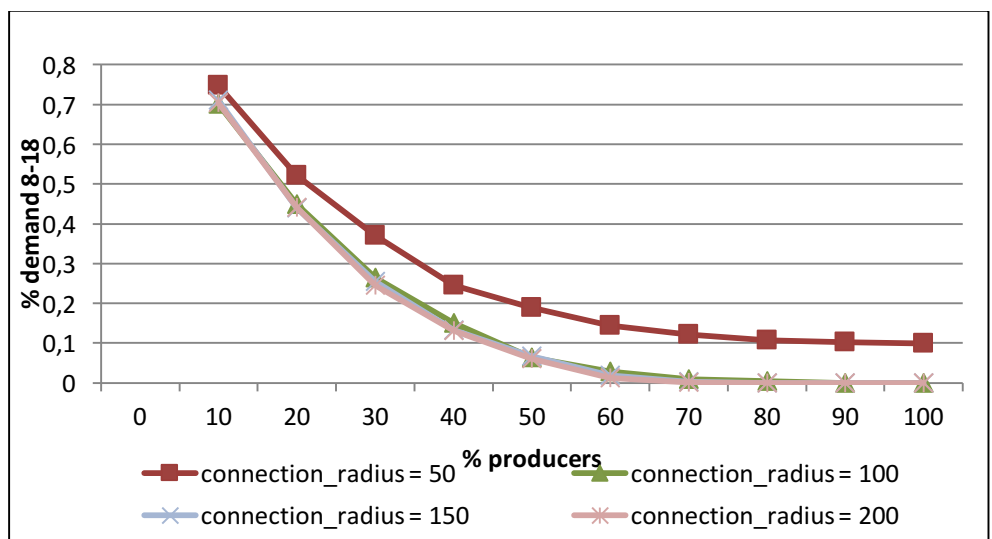

Fig.12 Trends of the *supply_percentage* index in correspondence to links activation threshold 0%

The $energy\_loss\_percentage$ index, approximately for each links activation threshold, shows an increasing trend. Actually, the value of the $connection\_radius$ equal to 50 m is responsible for a major (although limited) electricity loss. Nevertheless, beyond 50 m, the electricity losses are almost similar at the different distances. In addition, the increase of the percentage of producers does not avoid electricity loss; rather, the exceeding electricity is not distributed since neighbouring agents also produce electricity on their own and do not need to receive from the other producers.

The $supply\_percentage$ index exhibits an initial decrease and then performs a nearly constant trend. Generally, low percentages of producers mean a major amount of electricity supply from the central-agent in order to satisfy the demands of the agents of the network. However, the increase of the percentage of producers is a convenient choice up to the point, around the 60-70% of producers, beyond which no further advantages may be achieved (also because, for non-zero links activation thresholds, the supply becomes null). Therefore, although producers are able to reach a broader neighbourhood of agents, the supply from the central-agent does not decrease; this is due to the fact that agents have already distributed their exceeding electricity to closer neighbours and do not use further links for the distribution or they do not have further exceed to distribute.

Finally, the comparison of the $index_{mix}$ behaviour for different links activation thresholds by varying the radius of connection and the percentage of producers is shown in Fig.13.

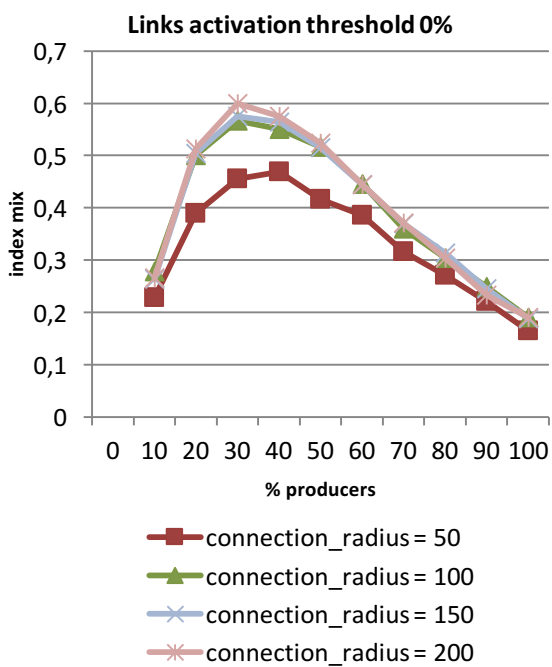
(a)

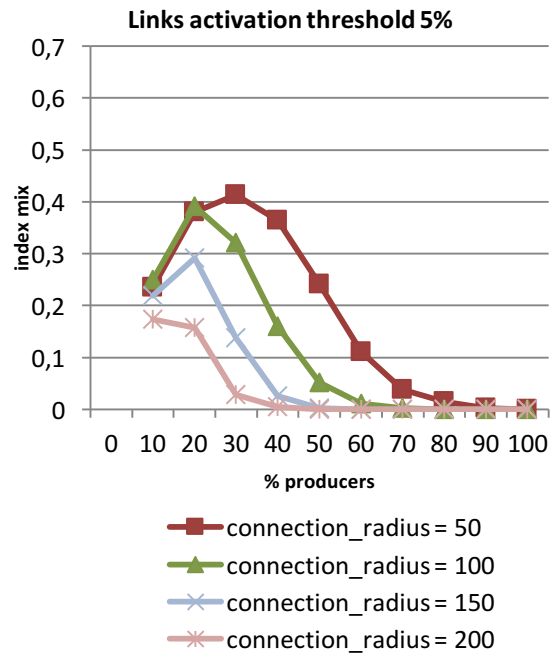
(b)

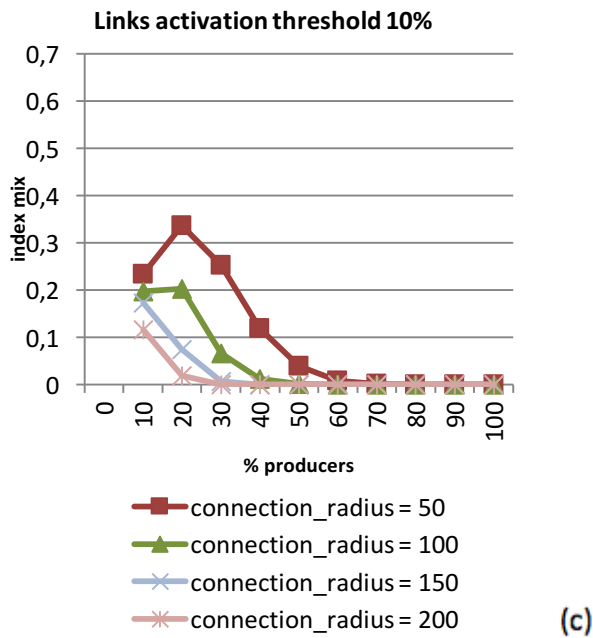

Fig.13 Trends of the $index_{mix}$ in correspondence to: (a) links activation threshold 0%; (b) links activation threshold 5%; (c) links activation threshold 10%

The results are qualitatively similar to the corresponding ones of scenario 1, shown in Fig.8; all curves show an initial increment and then a decrement that is more evident the higher becomes the considered links activation threshold. The best performance in terms of the $index_{mix}$ is generally attained in correspondence with the 30% of producers and distances higher than 50 m, when no links activation threshold is considered. However, the insertion of the thresholds diminishes the value of the $index_{mix}$; in these cases, the best trade-off among the three indexes is found for every configuration that plans the insertion of a percentage of producers equal to 30% and 20% (respectively for threshold 5% and 10%) and a $connection\_radius = 50\ m$.

The curves of the $index_{mix}$ for the different thresholds at fixed values of the $connection\_radius$ are reported in Fig.14.

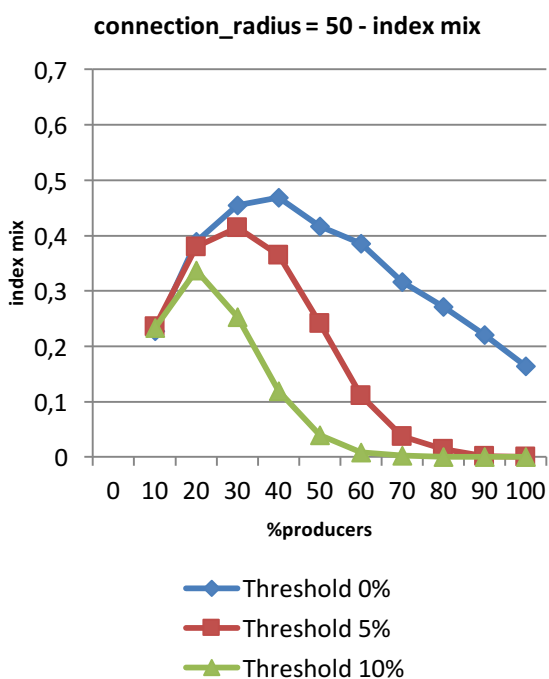
(a)

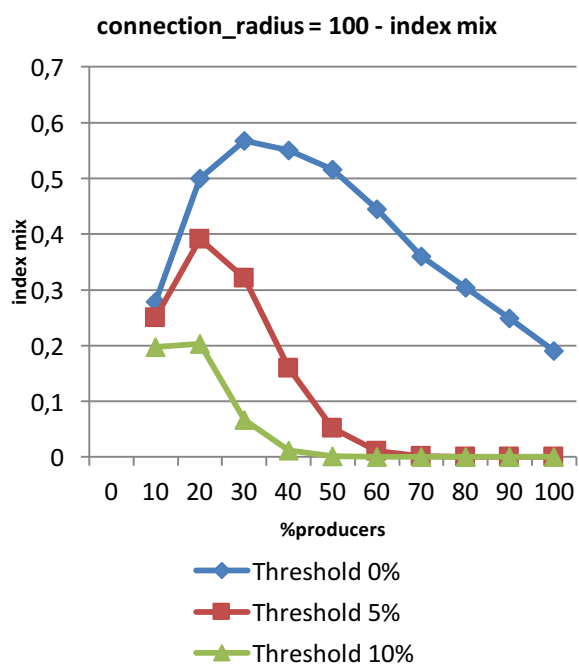
(b)

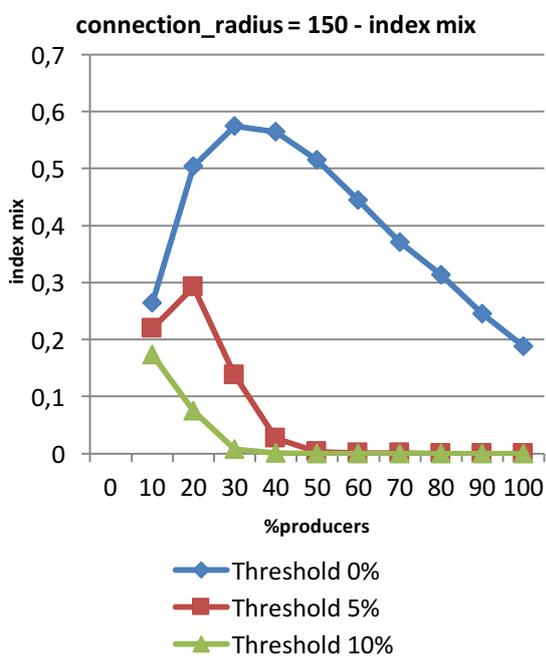 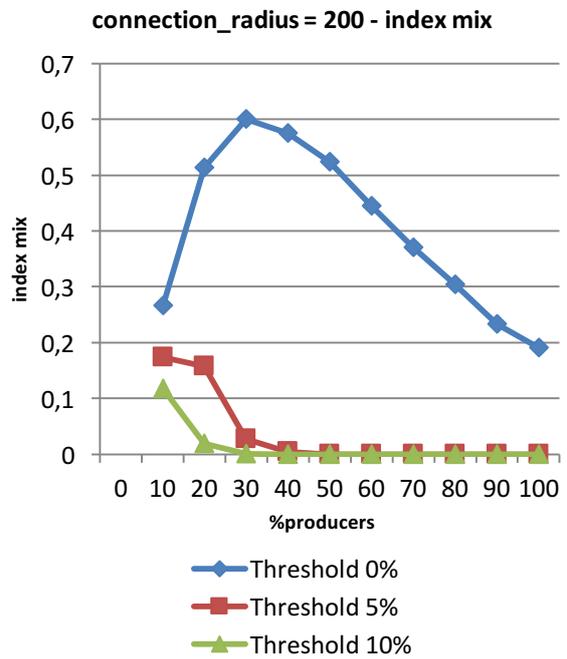

(c) (d)

Fig.14 Threshold comparison for the $index_{mix}$ at: (a) $connection\_radius = 50\ m$; (b) $connection\_radius = 100\ m$; (c) $connection\_radius = 150\ m$; (d) $connection\_radius = 200\ m$

As previously highlighted in the discussion of Fig.9 for scenario1, the thresholds produce a decrease of the recorded values of the $index_{mix}$. The decrease becomes more significant at increasing values of the $connection\_radius$. Hence, as also noticed, the higher the distance the lower the $index_{mix}$.

The comparison between scenario 1 and scenario 2 is briefly summarized in Fig.15. The $index_{mix}$ is reported in these graphs since it combines the major values of the links exploitation with the minor values of both the electricity losses and electricity supply from the central agent. To highlight the impact that the thresholds have on the evaluation of the performance of the distribution network, the sole 0% and 10% activation thresholds are compared and reported in the graphs.

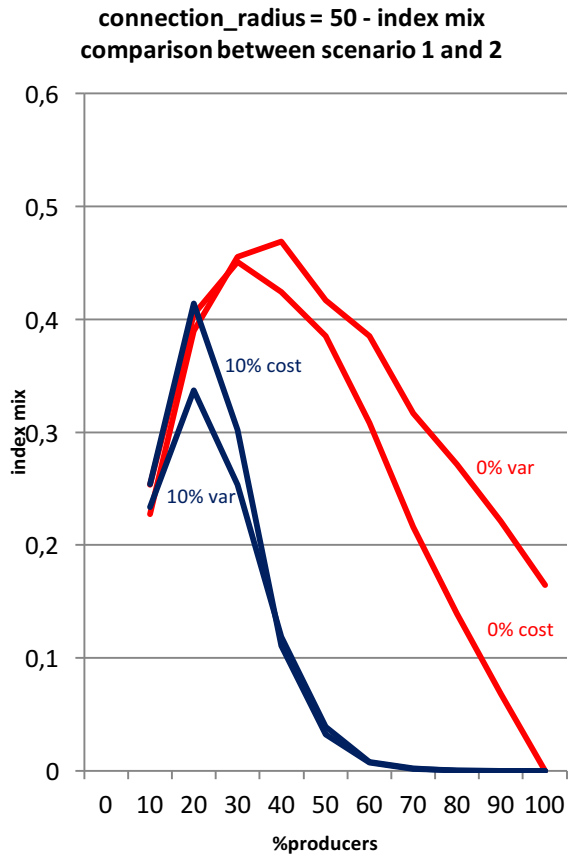
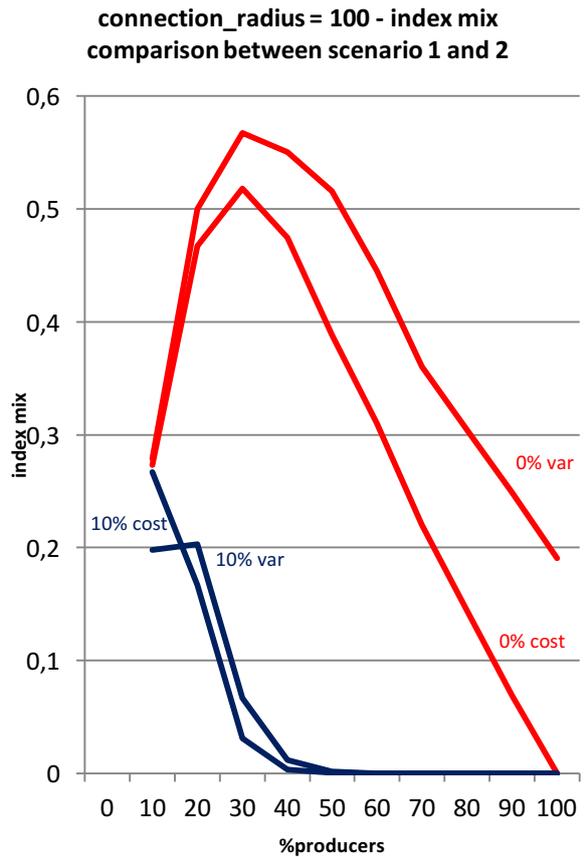
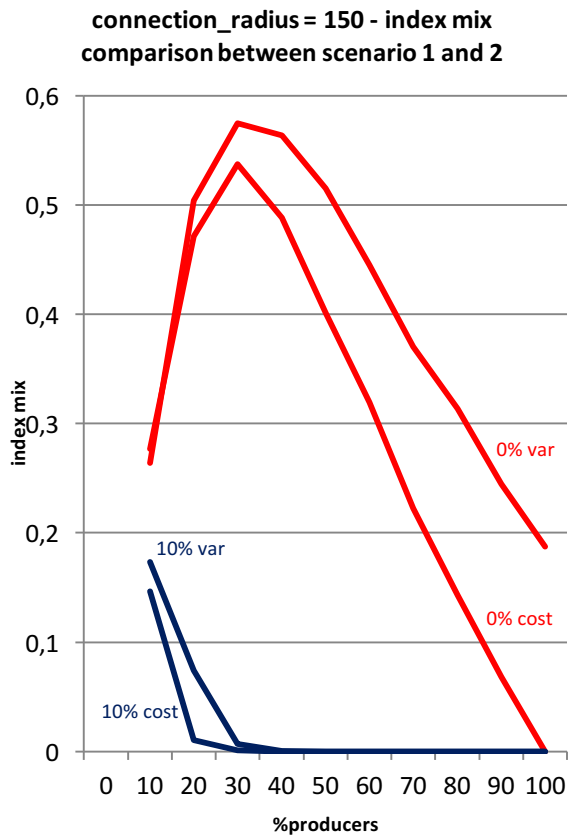
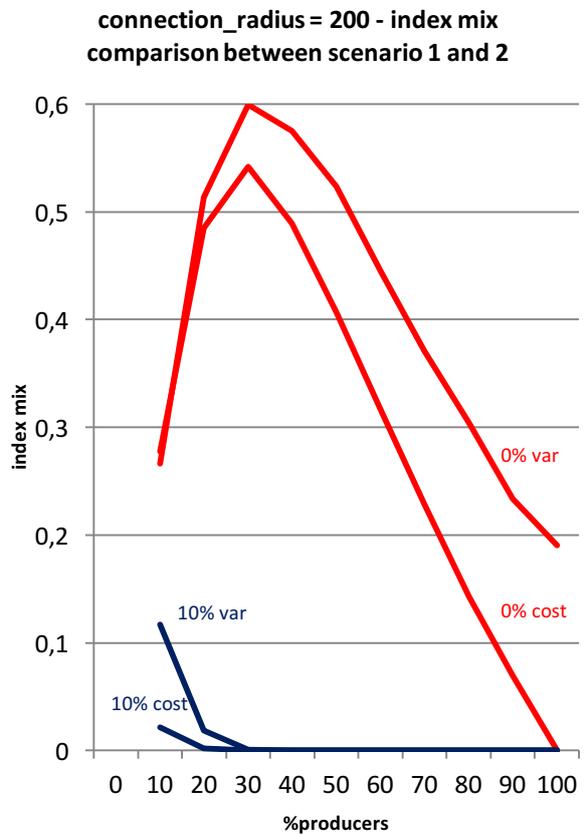

Fig.15 Threshold comparison for the $index_{mix}$ at: (a) $connection\_radius = 50\ m$; (b) $connection\_radius = 100\ m$; (c) $connection\_radius = 150\ m$; (d) $connection\_radius = 200\ m$

As a general observation, high distances combined with a nil threshold permit to achieve better performance of the distribution network, in terms of links exploitation and of both minor losses and central supply. As regard to the percentage of producers, instead, the 30% of producers among the totality of agents guarantees the maximum values of the $index_{mix}$ in every case. In details, a nil threshold means that every link that is used for at least one exchange during the operating time [08:00, 18:00] is included in the $links\_percentage$ index. Inserting the threshold of 10% is, however, a good argument to avoid fully unused links, since all links used less than the 10% of the time are excluded. Under this hypothesis, the best performances are recorded, oppositely, in correspondence with low distances, in particular for $connection\_radius = 50\ m$ and for the 20% of producers. Therefore, the planning of a network of energy distribution has also to consider the real exploitation of its links in order to avoid costly interventions that do not bring advantages in terms of energy distribution.

4. **Conclusions**

The insertion of renewable based energy systems within the urban areas allows citizens to shift from passive consumers to active producers, since they become able to both reach the energy self-sufficiency and distribute the eventual exceeding production. Considering consumers and producers as nodes and the energy exchanges as links, a network of energy distribution may be configured. The assessment of both the impact of renewable energy systems on the traditional supply and the configuration of the connections in the resulting network requires appropriate models. To the purpose, this paper introduces an agent-based model aimed at orienting the design process of an energy distribution network. The model introduces two breeds of agents; the node-agents, considered as consumers and potential producers, and the central-agent, representative of the traditional power plant. Simulations run to reproduce the daily exchange of energy among buildings in their capacity as consumers or producers equipped with renewable energy systems. Moreover, three indexes are introduced to describe the operation of the distribution network and refer to the rate of exploited links, the energy not further distributed and the energy supplied from the central-agent.

The model is tested for a theoretical area and two main scenarios are identified for the simulation process; on one side constant profiles of electricity demands and production of agents and, on the other side, variable profiles. Both scenarios consider percentages of agents equipped with renewable energy systems between 0% and 100% and four values of the admitted distance of distribution. Moreover, links activation thresholds are inserted in the model to evaluate how many time links are exploited within the total operating period.

The results of the simulations permit to conclude that the design of the distribution network within urban areas should consider the practical usability of the energy connections, beyond the evaluation of both the preferable distance of connection and percentages of producers. Indeed, from this point of view, the number of links used for the energy exchange strongly decreases when a constraint on their activation is posed. Therefore, planning long-distance connections may be considered as the best condition at a first glance, but, actually, the insertion of activation thresholds (such as the 10%) permits to conclude that the majority of the links is used less than the 10% of the time. Rather, low distances, typically around 50 m coupled with the 20-30% of producers, ensure best performances in terms of exploitation of the distribution network.

Further research will be oriented in the implementation of the presented model in a real case study, also including a cost analysis of the energy exchanges to define scenarios for the proper design of an urban energy distribution network.